\documentclass[graybox]{svmult}

\usepackage{type1cm}        
\usepackage{makeidx}         
\usepackage{graphicx}        
\usepackage{multicol}        
\usepackage[bottom]{footmisc}

\usepackage{newtxtext}       %
\usepackage[varvw]{newtxmath}       

\makeindex             

\usepackage{amsmath}
\DeclareMathOperator{\arctanh}{arctanh}

\def\be{\begin{equation}}
\def\ee{\end{equation}}

\def\bea{\begin{eqnarray}}
\def\eea{\end{eqnarray}}

\def\ba{\begin{array}}
\def\ea{\end{array}}

\newcommand*\diff{\mathrm{d}} 

\def\a{\alpha}
\def\b{\beta}

\def\e{\varepsilon}

\def\f{\frac}
\def\g{\gamma}

\def\l{\left}

\def\m{\mu}
\def\n{\nu}

\def\p{\partial}

\def\r{\right}

\def\z{\zeta}

\newcommand{\bb}{\,\square}

\begin{document}

\title*{Progress in Einstein-Cartan gravity}
\titlerunning{EC gravity}
\author{Mikhail Shaposhnikov}
\institute{Mikhail Shaposhnikov \at Institute of Physics,
\'Ecole Polytechnique F\'ed\'erale de Lausanne (EPFL),
CH-1015 Lausanne, Switzerland, \email{Mikhail.Shaposhnikov@epfl.ch}\\
\,\\
Contribution to the book ``Open Issues in Gravitation and Cosmology - Original Contributions, Essays and Recollections in Honour of Alexei Starobinsky'', to be published by Springer, edited by Andrei Barvinsky and Alexander Kamenshchik.
}
%
%
\maketitle


\abstract{It is well-known that the gravitational force can be obtained by gauging the Lorentz group, which puts gravity on the same footing as the Standard Model fields. The resulting theory - Einstein-Cartan gravity - has several crucial advantages. I will overview the construction of the Weyl-invariant version of this theory and discuss its applications in particle physics and cosmology, in particular for inflation and the strong CP problem.}

\section{Introduction}
\label{sec:intro}
There are several approaches to the theory of gravity. Historically the first one is that of Einstein, which tells that gravity is not an ordinary force, but rather a property of spacetime geometry. Thus, the theory of gravity is geometrodynamics (see e.g. the textbook \cite{Misner:1973prb}). Unifying it with the other forces of nature would then require geometrizing the photon and other fields of the Standard Model (SM). 

Yet another point of view (often attributed to Feynman \cite{Feynman:1963ax}) is that gravity is just like electrodynamics, but associated with a spin-2 massless particle - graviton, instead of the spin-1 massless photon. Then geometrodynamics is simply an effective low-energy theory. The Standard Model uses the gauging of the global symmetry group SU(3)xSU(2)xU(1), leading to the existence of the photon, intermediate vector bosons, and gluons. In full analogy with this construction, the gauging of the global Lorentz group \cite{Utiyama:1956sy, Kibble:1961ba, Sciama:1962} results in the existence of the graviton and thus gravitational interactions. The corresponding gauge fields are\footnote{General conventions:  Greek and capital Latin letters are used for spacetime and Lorentz indices, respectively. The signature of both the spacetime $g_{\a\b}$ and Minkowski $\eta_{AB}$ metrics is mostly plus.  We set $c=\hbar=1$.} the tetrad $e_\mu^A$ and the spin connection $\omega_\mu^{AB}$, leading to the Einstein-Cartan (EC) formulation of General Relativity (GR) ~\cite{Cartan:1922, Cartan:1923, Cartan:1924, Cartan:1925}. 	

Apriori, the EC formulation of gravity contains more degrees of freedom than the Einstein metric gravity (where the only dynamical variable is the metric $g_{\mu\nu}$), hidden in the 16 components of the tetrad field and the 24 components of the spin connection. However, many of these components happen to be non-dynamical, and different variants of the theory have just propagating graviton and at most two scalar fields. The presence of these scalar fields may lead to interesting consequences, such as cosmological inflation or a solution to the strong CP problem. The theory can be constrained further by imposing a Weyl symmetry, acting as pointwise rescalings of the metric (or of the tetrad), removing one scalar degree of freedom.

In this paper, we will provide a short overview of our works \cite{Shaposhnikov:2020aen, Shaposhnikov:2020gts, Shaposhnikov:2020frq, Karananas:2021zkl, Karananas:2021gco, Karananas:2023zgg, Karananas:2024xja} of the construction of EC gravity and its Weyl-invariant version.\footnote{These articles were heavily used for writing this contribution. Some new formulas and discussions are presented in Sections \ref{WIEC} and \ref{WT}.} One of our inspirations came from the Starobinsky theory of inflation, based on the purely geometric action in metric gravity, containing, in addition to the Einstein-Hilbert term, a term quadratic in the Ricci scalar. 

\section{The Starobinsky theory}
\label{sec:staro}
Pure gravity in the metric formulation is based on the Einstein-Hilbert action
\be
\label{EHact}
S_{\rm{EH}} = \frac{M_P^2}{2} \int \diff^4 x \sqrt{g}(L_0+L_2)\,, ~~L_0=\Lambda\,,~~L_2=R\,,
\ee
where $R$ is the Ricci scalar,  $\Lambda$  the cosmological constant, $M_P$  the Planck scale, and $g$  the determinant of the metric. It contains two physical degrees of freedom, those of the graviton.  The action (\ref{EHact}) includes two leading terms, $L_0$ and $L_2$, in $1/M_P^2$ expansion of a generic Lagrangian $L$ invariant under the full group of diffeomorphisms (Diffs).  The next-order terms represent operators of mass dimension four. These are:
\begin{equation}
\label{L4}
    L_4= \frac{1}{4f_R^2} R^2+C_1 W_{\mu\nu\lambda\rho}W^{\mu\nu\lambda\rho}+ C_2 E_4 +C_3 \Box R\,.
\end{equation}
Here $\Box=g^{\m\n}\nabla_\m \nabla_\n$ is the covariant D'Alembertian, and $f_R^2,~C_{1,2,3}$ are dimensionless constants, while
\begin{equation}
    E_4=W_{\mu\nu\lambda\rho}W^{\mu\nu\lambda\rho}+\frac{2}{3}R^2-2 R_{\mu\nu}R^{\mu\nu}\,,
\end{equation}
is the Euler density operator, and $W_{\mu\nu\lambda\rho}$ the Weyl tensor.  The terms $E_4$ and $ \Box R$ are total derivatives. Thus, they do not contribute to the equations of motion. The square of the Weyl tensor leads to the presence of extra degrees of freedom, some of them with the ``wrong'' sign of kinetic terms, resulting in instabilities and the presence of ``ghost'' particles, arguably leading to an inconsistent theory.\footnote{The presence of new degrees of freedom can make the theory renormalisable \cite{Stelle:1976gc}. } The unique theory including operators of mass dimension four which is free of ghosts is that with the action
\be
\label{Sact}
S_S=S_{\rm{EH}} + \frac{1}{4f_R^2} \int \diff^4 x \sqrt{g}R^2\,. 
\ee
In addition to the graviton, it contains just one extra spin zero degree of freedom - the ``scalaron''. The simplest way to see that is to introduce an auxiliary field $\chi$ (with the mass dimension one) that allows us to make the action linear in the scalar curvature (this trick will be used several times in what follows):
\be
\label{Sacteq}
S_S=S_{\rm{EH}} +\frac{1}{2}  \int \diff^4 x \sqrt{g}\left(R \chi^2 -\frac 1 2 f_R^2 \chi^4\right) \,.
\ee
The equations of motion for $\chi$ lead to $\chi^2=R/f_R^2$, and inserting this solution to  (\ref{Sacteq}) gives  (\ref{Sact}). Transformation of the action (\ref{Sacteq}) to the Einstein frame, with the help of conformal transformation,
\be
\label{conf}
g_{\mu\nu} \to \left(1+\frac{\chi^2}{M_P^2} \right)^{-1} g_{\mu\nu}\,
\ee
reveals that (\ref{Sact}) is nothing but 
\be
\int \diff^4 x \sqrt{g}\left[\frac{M_P^2}{2} R-\frac{1}{2}(\partial_\mu\phi)^2-\frac{f_R^2 M_P^4}{4}\left(1-\exp\left(-\sqrt{\frac{2}{3}}\frac{\phi}{M_P}\right)\right)^2\right]
\label{ST}
\ee
with
\be
\label{STcon}
\phi=\sqrt{\frac{3}{2}}M_P \log{\left(1+\frac{\chi^2}{M_P^2}\right)}\,.
\ee

The theory  (\ref{Sact}) is very interesting. It is written entirely in terms of geometrical quantities. It represents the very first theory of successful inflation - that of Starobinsky \cite{Starobinsky:1980te}, consistent with all current cosmological observations. 

\section{An overview of Einstein-Cartan gravitational theories }
\label{sec:1}

The Langrangian (\ref{Sact}) is based on the metric formulation of gravity, where the role of the dynamical variable is played by $g_{\mu\nu}$. This formulation is not unique and is not even a natural one when we need to describe fermions, which are an essential part of the Standard Model. A curved space action for fermions, in addition to Diff, is invariant under local Lorentz transformations. Thus, the fermion Lagrangian contains new (in comparison with the metric) geometric objects: the tetrad - $e^A_\mu$ and the gauge field of the local Lorentz group -  spin connection $\omega^{AB}_\mu$ (for a review, see ~\cite{Hehl:1976kj}). These variables are a priori independent (the tetrad and spin connection totalise 40 components, 16 and 24 respectively) and serve as a basis of the Einstein-Cartan (EC) formulation of gravity \cite{Cartan:1922, Cartan:1923, Cartan:1924, Cartan:1925}. The EC formulation fits well with the gauge principle used in the construction of the SM: to get the SM vector bosons, the global internal SU(3)xSU(2)xU(1) symmetry is gauged, whereas to get gravity the global Lorentz symmetry is gauged ~\cite{Utiyama:1956sy, Kibble:1961ba, Sciama:1962}. 

To build the EC gravity Lagrangian, the following structures are essential: the curvature 
\be
\label{eq:curv_def}
F_{\m\n}^{AB} = \p_\m \omega_\n^{AB} -\p_\n \omega_\m^{AB}+\omega^A_{\m C}\omega^{CB}_\n - \omega^A_{\n C}\omega^{CB}_\m \ , 
\ee
and the torsion
\be
\label{eq:tors_def}
T_{\m\n\rho}=e_{\mu A} T^A_{\n\rho}\ ,~~~T_{\m\n}^A = \p_\m e_\n^A -\p_\n e_\m^A+ \omega^A _{\m B}e^B_\n -\omega^A _{\n B}e^B_\m  \,. 
\ee
The torsion can be decomposed into irreducible vector  ($v_\mu,~\a_\mu$) and tensor ($\tau_{\m\n\rho}$) components as follows:
\be 
\label{eq:tors_all}
v_\m = T^\n_{~\m\n} \ ,~~~a^\m = E^{\m\n\rho\sigma}T_{\n\rho\sigma} \ ,~~~\tau_{\m\n\rho} =\frac 2 3 \l( T_{\m\n\rho} -v_{[\n} g_{\rho]\m} - T_{[\n\rho]\m} \r) \,,
\ee
where the square brackets mean antisymmetrisation with respect to a pair of indexes, and $ E^{\m\n\rho\sigma}=\e^{\m\n\rho\sigma}/\sqrt{g}$. There are  two scalar curvature invariants, that read
\be
\label{eq:f_def}
F \equiv \frac {1}{8} \epsilon_{ABCD}E^{\m\n\rho\sigma}F^{AB}_{\m\n}e^C_\rho e^D_\sigma \ ,~~~\text{and}~~~\tilde F \equiv E^{\m\n\rho\sigma}e_{\rho C}e_{\sigma D}F_{\m\n}^{CD}  \ .
\ee
The second structure is called the Holst term.

To find a relation between the EC formulation of gravity and the metric one the following formulas are helpful:
\begin{align}
\label{eq:f_decomp}
F & = \frac{R}{2} +\frac{1}{\sqrt{g}}\p_\m\l(\sqrt{g}v^\m\r) -\frac 1 3 v_\m v^\m + \frac{1}{48} a_\m a^\m +\frac 1 4 \tau_{\m\n\rho}\tau^{\m\n\rho}\ ,  \\
\label{eq:tilde_f_decomp}
\tilde F & = -\frac{1}{\sqrt{g}}\p_\m\l(\sqrt{g}
a^\m\r)+\frac 2 3 a_\m v^\m -\frac 1 2 \epsilon^{\m\n\rho\sigma}\tau_{\lambda\m\n}\tau^{\lambda}_{~\rho\sigma}\ , 
\end{align}
where the torsion-free Riemannian curvature (and the Ricci scalar $R$) is defined as usually in the metric gravity.

The analogue of (\ref{EHact}) in EC gravity, involving operators of dimension up to two, is
\begin{align}
\label{EC2}
S_{\rm{EC},2} = M_P^2  \int \diff^4 x \det{e} [F +&a  \tilde F+b v_\mu v^\mu +c a_\mu a^\mu + d a_\mu v^\mu  +\\
 &e \tau^{\mu\nu\rho} \tau_{\mu\nu\rho}+f E^{\mu\nu\rho\sigma}\tau^\lambda_{\mu\nu}\tau_{\lambda\rho\sigma}+\Lambda]\,,
\nonumber
\end{align}
where $a,b,...,f$ are dimensionless constants.  Despite the presence of many dynamical variables, this action is equivalent to (\ref{EHact}) and contains just a graviton. Indeed, with the use of (\ref{eq:f_decomp},\ref{eq:tilde_f_decomp}), one can see that the equations of motion for the torsion give $v_\mu=a_\mu=\tau_{\m\n\rho}=0$, leaving us with the Einstein-Hilbert action. 

The theories (\ref{EHact}) and (\ref{EC2}), linear in curvature, can be augmented by the fields of the Standard Model. In both cases, the non-minimal coupling of the Higgs field to the curvature invariants may give rise to cosmological inflation, driven by the Higgs field \cite{Bezrukov:2007ep}. In the first (metric) case, the predictions of Higgs inflation are very close to those of Starobinsky theory, whereas, for the EC gravity, they may substantially differ from them, in particular for the tensor-to-scalar ratio \cite{Shaposhnikov:2020gts}. The torsion remains non-dynamical, though the vectors $v_\mu$ and $a_\mu$ are not equal to zero and are expressed through the SM fields. With a specific choice of non-minimal couplings, the EC gravity can be converted \cite{Shaposhnikov:2020gts} to the metric gravity or the Palatini gravity \cite{Palatini:1919ffw}, and lead to Palatini Higgs inflation \cite{Bauer:2008zj}.

The analogue of (\ref{Sact}) in EC gravity is
\be
\label{EC0}
S_{\rm{EC}}=S_{\rm{EC},2}+S_{\rm{EC},4}\,,
\ee
\be
\label{EC4}
S_{\rm{EC},4}= \int \diff^4 x \det{e} \left[\f{1}{f^2} F^2 + \f{1}{\tilde f^2} \tilde F^2 + \f{2}{f_m^2} F \tilde F\right]\,.
\ee
It contains the squares of the scalar curvatures (\ref{eq:f_def}) and is free from ghosts.\footnote{In general, there are 10 quadratic invariants which can be constructed from the curvature. They are listed in~\cite{Diakonov:2011fs}. In addition, the dimension (mass)$^4$ operators include the fourth power of torsion, cross terms of curvature and torsion, as well as derivatives of torsion and curvature. If taken with arbitrary coefficients, they lead to the existence of ghosts and tachyons. Some specific choices of the coefficients lead to ghost-free theories at the level of linearised excitations above the flat background, see, e.g.  \cite{Karananas:2014pxa, Barker:2024juc}. However, non-linear effects may jeopardise this conclusion, see, e.g.  \cite{Barker:2023fem}.} The dimensionless couplings $f,~\tilde f$ and $f_m$ can be considered as the gauge couplings of the local Lorentz group.

It is easy to see that this theory contains two extra scalar degrees of freedom in addition to the graviton. Indeed, making the same trick as for Starobinsky's theory reveals that the quadratic in curvatures part of (\ref{EC4})  is equivalent to
\be
\label{decomp}
\f{1}{2f^2} F^2 + \f{1}{2\tilde f^2} \tilde F^2 + \f{1}{f_m^2} F \tilde F \to F \chi^2 + \tilde F a^2 -\f{1}{2} \alpha \chi^4-\f{1}{2}\beta a^4+\gamma\chi^2 a^2
\ee
with
\be
\f{1}{\alpha}=2 D f^2,~\f{1}{\beta}=2 D \tilde f^2,~\f{1}{\gamma}=2 D f_m^2,~D=\f{1}{f^2\tilde f^2}-\f{1}{f_m^4}\,,
\ee
where $\chi$ and $a$ are scalar auxiliary fields. As previously stated, the torsion appears algebraically, whereas the ``integrating out'' of it and the transformation of the theory to the Einstein frame shows that the fields $\chi$ and $a$ acquire the non-trivial kinetic terms. As well as the Starobinsky theory, the model (\ref{EC4}) with a certain choice of parameters can accommodate inflation. In addition, one of the scalar particles can be used as an axion, solving the strong CP problem (see below and \cite{Karananas:2024xja}).

\section{Scale invariant metric gravity}
\label{SImetric}
The theories discussed above contain an explicit mass scale - the Planck mass. An attractive class of theories are scale-free gravities, enjoying global scale invariance. They only include dimensionless coupling constants (giving hope of renormalisability), whereas the mass scale appears because of the spontaneous breakdown of the scale symmetry. 

The scale transformation acts on the metric, scalar ($\phi$), vector ($A_\mu$) and fermion ($\psi$)  fields as
\be
\label{scaleT}
g_{\mu\nu}\to q^{-2} g_{\mu\nu},~\phi \to q\phi,~A_\mu\to A_\mu,~ \psi\to q^{3/2} \psi\,,
\ee
respectively. Here $q$ is an arbitrary constant.

The Lagrangian of scale-invariant metric gravity is given by 
\be
\label{SI}
S_{SI}= \frac{1}{4f_R^2} \int \diff^4 x \sqrt{g}R^2\,. 
\ee
As previously, the square of the Ricci scalar can be removed with the help of an auxiliary field $\chi$. After the transformation to the Einstein frame, the action is
\be
\label{SIeq}
S_{SI}\to \frac{1}{2} \int \diff^4 x \sqrt{g}\left(M_P^2 R-  \frac{1}{2} f_R^2 M_P^4  - (\partial_\mu\tilde\chi)^2\right)  \,. 
\ee
where $\tilde\chi$ is the canonically normalised scalar field,  $\chi=M_P \exp[{\tilde\chi/(\sqrt{6}M_P)}]$. This is Einstein's gravity with a non-zero cosmological constant and a massless scalar field - dilaton, related to the spontaneous breaking of the scale invariance. If $f_R^2 >0$ $(f_R^2 <0)$, the theory admits a de Sitter (anti-de Sitter) background. Different aspects of quadratic gravity were considered in \cite{Alvarez-Gaume:2015rwa, Hell:2023mph, Karananas:2024hoh}. Around the flat Minkowski background, it represents a strongly coupled theory with unknown excitations \cite{Karananas:2024hoh}.

The theory (\ref{SI}) can be augmented by the fields of the Standard Model, most notably by the Higgs SU(2) doublet $H$. In the unitary gauge $H=(0,h/\sqrt{2})$, the relevant Lagrangian has the form
\be
L_H= \f{1}{2}\xi_h h^2 R - \frac{1}{2} (\partial_\mu h)^2-\f{\lambda h^4}{4}\,,
\ee
where $\lambda$ is the scalar self-coupling. After transformation to the Einstein frame the Lagrangian, besides the Einstein-Hilbert term, contains the scalar part
\be
\label{KT}
-\frac{1}{2} K_{ij}\partial_\mu\phi_i \partial_\mu\phi_j -V\,,
\ee
where $\phi_1=h,~~\phi_2=\chi$, and the kinetic matrix is
\be
\label{mix}
K_{ij}=\f{1}{2} \Omega\delta_{i1}\delta_{j1}+\f{3}{4} M_P^2 \f{\partial \log \Omega}{\partial\phi_i}\f{ \partial \log \Omega}{\partial\phi_j}\,,
\ee
where $ \Omega=M_P^2/(\chi^2 +\xi_h h^2)$. This is a metric of a two-sphere with scalar curvature $2/(3 M_P^2)$. The potential is given by
\be
\label{pot}
V=\f{1}{4}\Omega^2\left(f_R^2 \chi ^4+h^4 \lambda \right)\,.
\ee
The ground state of this theory automatically has a flat direction 
\be
\label{flat}
h=f_R \sqrt{\f {\xi_h}{\lambda}} \chi\,,
\ee
corresponding to a Goldstone massless degree of freedom.\footnote{Note that the massless dilaton does not lead to the fifth force and thus is harmless from the experimental point of view \cite{Wetterich:1987fm, Wetterich:1987fk,Shaposhnikov:2008xb,Ferreira:2016kxi}.} The vacuum energy is given by
\be
V_0=M_P^4\f {f_R^2\lambda}{4(\lambda+ f_R^2\xi_h^2)}\,,
\ee
and the mass of the Higgs scalar excitation is
\be
\label{MH}
m_H^2=M_P^2\f{f_R^2\xi_h(6\lambda+f_R^2\xi_h(1+6\xi_h))}{3(\lambda+ f_R^2\xi_h^2)}\,.
\ee
In the classical theory, both the vacuum energy and the Higgs mass go to zero simultaneously if $f_R \to 0$, implying that the smallness of the Higgs mass and of the cosmological constant may be related \cite{Karananas:2024xja}.\footnote{In the scale-invariant theory based on the linear in curvature action  
the dilaton must be introduced ``by hand'' \cite{Shaposhnikov:2008xb,Shaposhnikov:2008xi, Garcia-Bellido:2011kqb}. Its coupling to the Higgs field is not related to the cosmological constant.}  Taken at face value, this contradicts observations---the Higgs boson mass is too small if $f_R^2$ is chosen to fit the observed value of the cosmological constant, whereas the cosmological constant is too large if  $f_R^2$ is fixed by comparing (\ref{MH}) with the observed Higgs mass. However, the classical action is not the whole story. The smallness or absence of the Higgs mass persists in scale-invariant quantum perturbation theory, as the zero values of these parameters correspond to fixed points of their renormalization group evolution~\cite{tHooft:1973mfk, Wetterich:1983bi}. This makes it in principle computable in terms of the parameters of the underlying theory, once nonperturbative effects are taken into account \cite {Coleman:1973jx, Weinberg:1976pe, Linde:1977mm, Shaposhnikov:2018xkv, Shaposhnikov:2018jag, Shaposhnikov:2020geh}.  A possibility certainly worth exploring is that the hierarchy between the Fermi and Planck scales be attributed to the semiclassical suppression of gravitational-Higgs instanton amplitudes, as proposed in~\cite{Shaposhnikov:2018xkv, Shaposhnikov:2020geh} (see also~\cite{Shaposhnikov:2018jag, Karananas:2020qkp}).

\section{Scale invariant EC gravity}
\label{SIEC}
Let us see now what happens in the scale-invariant pure EC gravity. The scale transformations of the gravitational degrees of freedom are
\be
\label{eq:trans_fields}
 e^A_\m~\mapsto~q^{-1} e^A_\m \ ,~~~\omega_\m^{AB}~\mapsto~\omega_\m^{AB} \,.
\ee
This leads to the following transformation laws for the Ricci scalar and the irreducible components of torsion
\be
\label{eq:trans_obj}
 R~\mapsto~q^2 R\ ,~~~v_\m~\mapsto~v_\m \ , ~~~a_\m~\mapsto~a_\m \ ,~~~\tau_{\m\n\rho}~\mapsto~q^{-2}\tau_{\m\n\rho} \ .
\ee
The transformations of the matter fields are given by (\ref{scaleT}).

A ghost-free scale-invariant EC action is given by $S_{\rm{EC},4}$ defined in eq. (\ref{EC4}).  At first sight, the resulting theory contains two extra degrees of freedom, associated with the auxiliary fields $\chi$ and $a$. This happens to not be the case.  A peculiar feature of the action $S_{\rm{EC},4}$ is that in addition to the global scale symmetry, it is invariant under a wider group, namely local Weyl transformations. This symmetry acts on the gravity variables as follows: eq. (\ref{eq:trans_fields}) remains in force, but now with $q$ being an arbitrary function of space-time coordinates. Also, the transformations of $a_\m$ and $\tau_{\m\n\rho}$ in (\ref{eq:trans_obj}) are the same, whereas the scalar curvature and the torsion vector $v_\m$ transform as\footnote{The fact that the torsion vector $v_\mu$ may be used as the Weyl gauge field was pointed out in \cite{Nieh:1981xk, Dereli:1982xb, Obukhov:1982zn}.}
\bea
\label{eq:Weyl_trans_R}
& &  R~\mapsto~q^2 R + 6 q \square q -12 \l(\p_\m q\r)^2 \ ,\\
\label{eq:Weyl_trans_v}
& & v_\m~\mapsto~v_\m +3q^{-1}\p_\m q \ .
\eea
 As a result, the theory contains just one extra degree of freedom: one can make a Weyl transformation removing one of the scalar fields (say, $\chi$), or, equivalently choose the gauge $\chi=M_P$. The action has the form
\be
S_{\rm{EC},4}\to \int \diff^4 x \sqrt{g}\left( \frac{1}{2}M_P^2 R  -  \frac{1}{2}(\partial_\mu\tilde a)^2 -V(\tilde a) \right) \,,
\ee 
with the potential given by \cite{Karananas:2025xcv}
\be
\label{potpure}
\frac{1}{32} M_P^4   \left(16 \alpha +\beta    \sinh^2\left({\sqrt{2/3} \tilde a/M_P}\right)
-8\gamma  \sinh\left({\sqrt{2/3} \tilde a/M_P}\right)\right)\,,
\ee
where $\tilde a=\sqrt{\frac{3}{2}} M_P  \arctanh\left(\frac{4a^2}{\sqrt{16a^4+\text{MP}^4}}\right)$ is the canonically normalised scalar field. The choice $\alpha=\gamma^2/\beta$ nullifies the cosmological constant in the minimum of the potential, while for $\gamma\gg\beta$ the theory leads to inflation with predictions close to Starobinski and Higgs inflations  \cite{Karananas:2025xcv}.

One can add the Higgs field $h$ to the pure gravity theory. The scale-invariant action (in the unitary gauge) has the form
\be
\label{ECSI}
S_{ECSI}=S_{ECWI}+S_{ECWB}\,,
\ee
where $S_{ECWI}$ is the Weyl invariant action for gravity and scalar field,
\be
\label{ECWI}
S_{ECWI}=S_{\rm{EC},4} +S_{\rm{HWI}}\,,
\ee
\be
\label{HWI}
S_{\rm{HWI}}
= - \f{1}{2} (D_\mu h)^2-\f{\lambda}{4}h^4 +h^2\left(\xi_h  F +\zeta_h  \tilde F+ c_a a_\mu^2\right)
\ee
with 
\be
\label{weylder}
D_\mu = \partial_\mu+\f{1}{3}v_\mu
\ee
being the Weyl covariant derivative. The Weyl invariance is broken explicitly by the scale-invariant term $S_{ECWB}$, given by
\be
\label{ECWB}
S_{ECWB}=h^2 \left(c_v v_\mu^2+ c_{av} v_\mu a_\mu\right)+\partial_\mu (h^2)\left(\zeta_{vh} v_\mu+\zeta_{ah} a_\mu\right)\,.
\ee
In addition to the graviton and the Higgs boson, this theory has two extra degrees of freedom, related to the auxiliary fields $\chi$ and $a$. The theory (\ref{ECSI}) is unlikely to have an acceptable low-energy phenomenology. Indeed, the Weyl invariance is restored in the limit of the vanishing Higgs field, leading to $S_{ECWB}\to 0$. This means that the kinetic matrix for the three fields $h,~a$ and $\chi$ (analogue of $K_{ij}$ in (\ref{KT})) has zero eigenvalue when $h\to 0$ (it can be shown that the eigenstate associated with it is $\chi+a$) .  Therefore, the Higgs boson interacts strongly with the massless dilaton - a Goldstone boson of the spontaneously broken scale invariance, potentially contradicting experimental constraints on the decay rate of the Higgs boson into invisible particles. 

\section{Weyl invariant gravities}
\label{WIEC}
A natural generalisation of the global scale invariance is the local, Weyl symmetry. In the metric formulation of gravity, it acts on the fields as in eq. (\ref{scaleT}), but now with $q$ being an arbitrary function of a space-time point. A set of transformation rules of different geometric quantities can be found in \cite{Nakayama:2013is},  see eq. (\ref{eq:Weyl_trans_R}) for the transformation of the scalar curvature. In the pure gravity case, the unique Weyl-invariant action is described by a square of the Weyl tensor $W_{\mu\nu\lambda\rho}$. The theory contains ghosts at the classical level, and whether it makes sense quantum mechanically is a subject of debate (see, e.g. \cite{Holdom:2024onr, Salvio:2014soa}). It also suffers from the Weyl anomaly \cite{Duff:1993wm}, leading to its breakdown to global scale symmetry \cite{Shaposhnikov:2022zhj, Shaposhnikov:2022dou}.

Going to the EC formulation of gravity changes the situation\footnote{Yet another attractive possibility for construction of Weyl-invariant gravities is associated with Weyl geometry \cite{Ghilencea:2023sti}. It has several similarities with the EC gravity discussed here, but there are also important differences.}. Contrary to the metric case, the curvature given by eq. (\ref{eq:curv_def}) is Weyl covariant.  The torsion tensor $\tau_{\mu\nu\rho}$ and tetrad $e_\mu^A$ transform in a uniform way, axial torsion vector $a_\mu$ does not change, whereas the vector torsion transforms as a gauge field corresponding to the local Weyl symmetry, eq. (\ref{eq:Weyl_trans_v}). The latter fact allows adding matter fields in a Weyl-invariant way,  by constructing a covariant derivative, as was done in eq. (\ref{weylder}). 

Thus, a pure EC gravity theory with the action (\ref{EC4}) is a healthy Weyl-invariant theory. The Standard Model can be added to it in a Weyl-invariant way. The Weyl-invariant Higgs action is that of eq. (\ref{HWI}), where the Weyl covariant derivative should be augmented by the usual SU(2)xU(1) terms. The Standard  SU(3)xSU(2)xU(1) kinetic terms for the gauge fields are Weyl invariant. The Weyl-invariant fermionic action $S_{\rm f}$ (written for one generation) is
\be
\label{eq:fermionic_action}
S_{\rm f} = \int \diff^4 x\sqrt{g} \Bigg[ \f i 2 \overline{\Psi}\g^\m D_\m\Psi  + \text{h.c.} + \l(\z^a_V V_\m +z^a_A A_\m\r)a^\m \Bigg] \ ,
\ee
where $\z^a_V,z^a_A$ are real constants, and 
\be
V_\m = \bar{\Psi} \gamma_\mu \Psi \ ,~~~A_\m = \bar{\Psi} \gamma_5 \gamma_\mu \Psi \ ,
\ee
are the vector and axial fermionic currents, respectively.  The fermionic covariant derivative $D_\m$ is defined as
\be
\label{eq:ferm_covariant_der}
D_\m = \mathcal D_\m +\frac 1 8 \omega_\m^{~AB}\left(\g_A\g_B - \g_B\g_A\right) \ ,
\ee
with $\mathcal D_\m$  the appropriate (flat spacetime) SM covariant derivative, the exact form of which depends on the specific nature of the fermion (left/right handed, lepton/quark). 
Yet another admitted Weyl invariant term is the coupling of the Higgs hypercharge current with the axial torsion vector,
\be
\label{Zint}
\zeta_Z Im  [H^\dagger \mathcal D_\m H] a_\m\,,
\ee
where $\zeta_Z $ is some real constant.

In addition to the SM fields and the graviton, this theory has just one extra degree of freedom. In the gauge $\chi=M_P$ and after transformation to the Einstein frame it can be identified with the auxiliary field $a$. Since The Weyl symmetry is local, there is no massless Goldstone scalar. The spectrum is massive with the exception of the graviton and the photon.  

The combination of the Weyl-invariant gravity and the SM  has an interesting low-energy phenomenology, allowing to unify the strong CP and hierarchy puzzles \cite{Karananas:2024xja}. If the Lorentz gauge couplings $f,~\tilde f$ and $f_m$ (or, what is the same, $\alpha,~\beta$ and $\gamma$) are tiny and of the same order of magnitude, the tree, classical masses of the Higgs boson, the field $a$, and the cosmological constant are also small, as all these quantities are proportional to them. The minimum of the potential is at 
\be
\label{min}
h^2=2 \xi_h M_P^2\frac{\alpha\beta-\gamma^2}{\lambda\beta},~~a^2=M_P^2\frac{\gamma}{\beta}\,,
\ee
leading to the vacuum energy
\be
\label{evac}
\epsilon_{vac}=M_P^4 \frac{\alpha\beta-\gamma^2}{2\beta}\,,
\ee
and the masses of the excitations are
\be
\label{masses}
m_h^2=4 \xi_h M_P^2 \frac{\alpha\beta-\gamma^2}{\beta}\,,~~
m_a^2= M_P^2 \frac{\beta^2+16\gamma^2}{24\beta}\,.
\ee
The discussion after eq. (\ref{MH}) is fully applicable to this case, leading to a potential computability of the Higgs boson mass. 

The field $a$ -- an axion-like particle (ALP)  has a non-trivial coupling to the axial and vector fermionic currents, and has all the requisites to solve the strong CP-problem, provided its ``gravitational'' mass is small compared with the QCD induced mass \cite{Karananas:2024xja}.  

In addition, this theory is fully compatible with the neutrino Minimal Standard Model ($\nu$MSM)~\cite{Asaka:2005an, Asaka:2005pn}, which is a minimal extension of the SM in the neutrino sector capable of addressing simultaneously the experimental problems of the latter: neutrino masses and oscillations, dark matter, baryon asymmetry of the Universe. Weyl symmetry forces the tree-level Majorana masses of the heavy neutral leptons (HNLs) of the $\nu$MSM to be zero, which would be incompatible with phenomenology since successful baryogenesis cannot take place. In full analogy to the situation with the Higgs, one can speculate that non-perturbative effects generate masses for the HNLs. This is in line with the common lore that gravity breaks all global symmetries (see e.g.~\cite{Kallosh:1995hi}). In this case, the classical action would have a global $B-L$ symmetry in the absence of HNL masses ($B$ and $L$ are the baryon and lepton numbers, respectively), and the breaking of $B-L$ \`a la Nambu-Jona-Lasinio~\cite {Nambu:1961tp, Nambu:1961fr} can potentially lead to Majorana masses for the HNLs, even though the order of magnitude of this effect remains obscure and has never been computed. As an additional bonus, the EC formulation of GR provides a mechanism for generating the HNLs in the early Universe so that the lightest of them can provide the observed abundance of dark matter in a wide range of masses~\cite{Shaposhnikov:2020aen}.

\section{From classical to quantum theory}
\label{WT}
The transition from classical to quantum Weyl-invariant theories is not straightforward for several reasons. 

First,  all realistic {\em renormalisable}  conformally invariant at the classical level field theories\footnote{Weyl invariance implies conformal invariance on the top of the flat Minkowski metric.},  suffer from the scale anomaly: the divergence of the dilatational current is non-zero due to quantum effects and is proportional to the $\beta$-functions of dimensionless couplings, governing their renormalisation group running \cite{Coleman:1970je}. 

Second,  {\em in the metric formulation of gravity}  the Weyl anomaly (for a review see \cite{Duff:1993wm})  forbids keeping the classical Weyl symmetry in a quantum theory in a non-trivial gravitational background metric.

The solution to the first problem was suggested almost 50 years ago in \cite{Englert:1976ep} (see also \cite{Shaposhnikov:2008xi}). The reason for the presence of quantum scale anomalies is connected to the fact that any regularisation of divergent Feynman graphs of renormalisable field theories contains an explicit mass scale. It can be an ultraviolet (UV) cutoff $\Lambda$ or mass $M_{PV}$ in Pauli-Villars regularisation, or the scale $\mu$ in dimensional regularisation (DimReg), eliminating a mismatch between coupling constants in different dimensions. These scales break  conformal invariance explicitly, resulting in the conformal anomaly. The idea of \cite{Englert:1976ep}, who used  DimReg, consists in replacing $\mu$ by a dynamical field - dilaton $\chi$. This makes the theory conformally invariant in $D=4-2\epsilon$ dimensions and allows the subtraction of divergencies in a conformally invariant way. The price to pay is the renormalisability of the theory: the Lagrangian in D dimensions contains fractional powers of the dilaton field, leading to the proliferation of different evanescent operators needed to remove the divergencies \cite{Shaposhnikov:2008xi, Shaposhnikov:2009nk}. The spontaneous breaking of the Weyl invariance -- the non-zero dilaton vev -- is automatically embedded in the formalism. Of course, the use of DimReg for the construction of Weyl invariant theories is not unique, everything works with any type of regularisation: simply replace the cutoff  \cite{Wetterich:1987fm, Wetterich:1987fk}, lattice spacing \cite{Shaposhnikov:2008ar} or the Pauli-Villars mass with the dynamical dilaton field. As usual, the DimReg is more suited to practical computations, which in this context can run up to several loops \cite{Ghilencea:2016ckm}.

The solution to the second problem is the use of another gravity formulation, such as Einstein-Cartan gravity, or the Weyl geometry instead of the Riemann one \cite{Ghilencea:2023sti}.  

In the metric formulation of gravity, the reason for the Weyl anomaly \cite{Duff:1993wm} can be traced to the impossibility \cite{Komargodski:2011vj, Luty:2012ww, Shaposhnikov:2022dou, Shaposhnikov:2022zhj} to have local\footnote{However, {\em non-local} operators providing the Weyl invariant action can be constructed \cite{Fradkin:1983tg, Antoniadis:1991fa}.} Weyl and Diff invariant generalisation of the following conformally invariant action in {\em four dimensions} ($D=4$)\footnote{We define the ``conformally'' invariant action as that written in flat space-time and invariant under conformal group including the Poincare transformations, scale transformation, and special conformal transformations.}
\begin{equation}
\label{4d conformal}
S_\tau=\int d^4x (\bb\tau)^2\,,
\end{equation}
where $\tau=\log(\phi/\mu)$ and $\phi$ is the scalar field with mass dimension one\footnote{For the scalar field $\tau$ with the mass dimension zero the action $\int d^4x(\bb \tau)^2$ allows a Weyl and Diff invariant generalisation with the help of the Riegert operator \cite{Fradkin:1981jc, Paneitz:2008afy, Riegert:1984kt}.}. The term $\tau \Box^2\tau$ will inevitably appear in an effective theory for the dilaton \cite{Komargodski:2011vj}  in spontaneously broken conformally invariant theory in flat space-time.  

If $D\neq 4$, such a generalisation exists \cite{Shaposhnikov:2022zhj}, namely
\begin{equation}
\label{div}
S_\tau \to \lim_{D\to 4} \int d^D x \sqrt{-g}\left[\tau\Delta_4\tau+
2\tau \left(-\frac{1}{6} \bb R +\frac{1}{4} E_4\right)+\frac{R^2}{36} + L_{anom}\right]~, 
\end{equation}
where  $\Delta_4$ is the Riegert operator \cite{Fradkin:1981jc, Paneitz:2008afy, Riegert:1984kt},
\begin{equation}
 \Delta_4=\bb^2+2R^{\mu\nu}\nabla_\mu\nabla_\nu-\frac{2}{3}R\bb+\frac{1}{3}(\nabla^\mu R) \nabla_\mu
\end{equation}
and
\begin{equation}
\label{anom}
 L_{anom}=\frac{E_4}{2(D-4)}~.
\end{equation}
The Euler density $E_4$ is a total derivative only in $D=4$ so it cannot be dropped if $D\neq 4$, leading to an ill defined expression at  $D=4$, i.e. to Weyl anomaly.

In the EC gravity, where the vector torsion plays the role of the Weyl gauge field, every derivative can be promoted to the Weyl covariant one as in eq. (\ref{weylder}), removing the obstacle for Weyl invariance existing in the metric formulation of gravity (a similar argument applies to the Weyl geometry \cite{Ghilencea:2023sti}).

For quantum computations, the EC theory can be formally continued into $D=4-2\epsilon$ space-time, to use dimensional regularisation. The curvature invariants, transforming uniformly under Weyl transformations as $F\to q^2 F,~\tilde F\to q^2 \tilde F$ are written in D-dimensions as 
\begin{align}
\label{eq:f_decompD}
F & \to \frac{R}{2} +\frac{1}{\sqrt{g}}\p_\m\l(\sqrt{g}v^\m\r) -\frac{D-2}{2(D-1)} v_\m v^\m + \frac{1}{48} a_\m a^\m \ ,  \\
\label{eq:tilde_f_decompD}
\tilde F & \to  -\frac{1}{\sqrt{g}}\p_\m\l(\sqrt{g}
a^\m\r)+\frac {D-2}{D-1} a_\m v^\m \ , 
\end{align}
where we omitted the contribution of the tensor torsion since equations of motion nullify it (this allows us to ignore the well-known problems with the definition of the Levi-Civita tensor in DimReg). The transformation of the tetrad determinant is $det[e]\to q^{-D} det[e]$, and that of the vector torsion is
\be
\label{vecD}
v_\m~\mapsto~v_\m +(D-1)q^{-1}\p_\m q \,.
\ee
The covariant derivative of the Higgs field (we keep its mass dimension equal to one independently of $D$) is given by 
\be
\label{weylderD}
D_\mu = \partial_\mu+\f{1}{D-1}v_\mu \,.
\ee
A construction of the Weyl-invariant action in $D$ dimension is not unique. Namely, any Weyl covariant term in the pure gravity action
(\ref{decomp}) (e.g. $F\chi^2$), in the Higgs action (\ref{HWI}) (e.g. $F h^2$) , fermion action (\ref{eq:fermionic_action}), etc can be multiplied by $\chi^{D-4}\rho(a/\chi)$, where $\rho$ is an arbitrary function with the property $\rho\to1$ when $D\to4$. For instance, following \cite{Shaposhnikov:2008xi}, the pure gravity action can be written as 
\be
\label{DWeyl}
S^D_{\rm{EC}}=
\f{1}{2} \int \diff^D x \det{e} \left[F \chi^{D-2} + \tilde F a^{D-2} -\f{1}{2} \alpha \chi^D-\f{1}{2}\beta a^D+\gamma\chi^{D/2} a^{D/2}\right]\,,
\ee
where $\alpha,\beta$ and $\gamma$ are now the functions of the ratio $a/\chi$ and $D$ with the property $\alpha,\beta,\gamma\to const$ when $D\to4$.

All these actions give the same four-dimensional action. They lead, however, to different quantum Weyl-invariant physics because of evanescent terms related to UV divergences producing poles in $\epsilon$. In particular, the different prescriptions on how to handle the logarithmic corrections to the effective action originate from this arbitrariness \cite{Bezrukov:2009db, Shaposhnikov:2018nnm}.

\section{Conclusions}
\label{sec:5}
The use of the Einstein-Cartan formulation of gravity may provide several benefits:
\begin{itemize}
\item It allows the construction of anomaly-free quantum Weyl-invariant theories of particle physics and gravity.
\item The combination of the Standard Model with the Weyl-invariant Einstein-Cartan gravity automatically contains just one extra scalar degree of freedom (ALP) with all properties to solve the strong CP problem.
\item The smallness of the cosmological constant, the ALP, Higgs boson, and heavy neutral lepton masses in the $\nu$MSM results from tiny values of the dimensionless gauge couplings in Einstein-Cartan gravity, which opens the possibility of potentially computing these parameters from non-perturbative
effects. 
\end{itemize}

Of course, there are many open problems to be solved. From the theoretical side, the quantum theory of the Weyl-invariant Einstein-Cartan gravity remains to be developed, with an understanding of its high-energy limit and non-perturbative effects, which may lead to the computation of the Higgs boson and heavy neutral lepton masses. This theory is not perturbatively renormalizable. We may think about its ultraviolet completion along the lines of asymptotic safety \cite{Weinberg:1980, Reuter:1996cp, Berges:2000ew}, classicalization~\cite{Dvali:2010bf, Dvali:2010jz, Dvali:2011th}, or non-renormalizable resummation of amplitudes proposed in~\cite {Shaposhnikov:2023hrg}. From the phenomenological side, the most interesting questions are related to cosmology. It would be important to understand whether inflation can take place in this theory and what would be the predictions of observables. The presence of a new light scalar field poses the question of whether this can be a suitable dark matter candidate. These questions are now under investigation \cite{ToAppear}.

\begin{acknowledgement}
I thank my collaborators  Georgios Karananas,  Andrey Shkerin,  Inar Timiryasov, Anna Tokareva, Sebastian Zell and Daniel Zenhausern for developing together the ideas described in this contribution. The comments of Georgios Karananas and Sebastian Zell on the manuscript are greatly appreciated. This work was supported in part by the Generalitat Valenciana grant PROMETEO/2021/083.
\end{acknowledgement}

 \bibliographystyle{utphys}
\providecommand{\href}[2]{#2}\begingroup\raggedright\endgroup

\end{document}